\documentclass[reprint, amsmath, amssymb, aps, pra, floatfix, twocolumn, showkeys]{revtex4-2}

\usepackage[caption=false]{subfig}
\usepackage{cleveref}
\usepackage{amsmath}
\usepackage{graphicx}

\renewcommand{\eqref}[1]{eq.~\!(\ref{#1})} 
\newcommand{\Eqref}[1]{Eq.~\!(\ref{#1})} 

\newcommand{\figref}[1]{figure~\!\ref{#1}}

\newcommand{\tr}[1]{ \text{tr} \! \left({#1}\right) }
\newcommand{\dd}{\,\text{d} }
\newcommand{\grad}{\nabla  }

\newcommand{\eeq}{=}
\newcommand{\Cos}[1]{\cos\!\left(#1\right)}
\newcommand{\Sin}[1]{\sin\!\left(#1\right)}
\newcommand{\Cot}[1]{\cot\!\left(#1\right)}
\newcommand{\Csc}[1]{\csc\!\left(#1\right)}

\newcommand{\Tan}[1]{\tan\!\left(#1\right)}

\newcommand\bigo[1]{\mathcal{O}\!\left(#1\right)}

\newcommand{\Diffu}{\mathcal{D}} 

\usepackage{enumerate}

\usepackage{pgffor}
\usepackage{pdfpages}
\makeatletter
\AtBeginDocument{\let\LS@rot\@undefined}
\makeatother

\begin{document}

\preprint{APS/123-QED}

\title{A continuum limit for dense spatial networks}

\author{Sidney Holden$^{1,2}$}\email{sholden@flatironinstitute.org}
\author{Geoffrey Vasil$^{2}$}\email{gvasil@ed.ac.uk}

\affiliation{$^{1}$Center for Computational Biology, Flatiron Institute, New York, NY, USA 10010}
\affiliation{$^{2}$School of Mathematics and Maxwell Institute for Mathematical Sciences, The University of Edinburgh, EH9 3FD, United Kingdom}

\begin{abstract}
Many physical systems---such as optical waveguide lattices and dense neuronal or vascular networks---can be modeled by \textsl{metric graphs}, where slender ``wires'' (edges) support wave or diffusion equations subject to Kirchhoff conditions at the nodes. This work proposes a continuum-limit framework that replaces edge-based equations with a global coarse-grained partial differential equation (PDE) defined on the continuous space occupied by the network. The derivation naturally introduces an \textsl{edge-conductivity tensor}, an \textsl{edge-capacity function}, and a \textsl{vertex number density} to encode how each microscopic patch of the graph contributes to the macroscopic phenomena. The results have interesting similarities and differences with the Riemannian Laplace-Beltrami operator. We calculate all macroscopic parameters from first principles via a systematic discrete-to-continuous local homogenization, finding an anomalous effective embedding dimension resulting from a homogenized diffusivity. Numerical examples---including an axisymmetric “spiderweb”, several periodic lattices, random Delaunay triangulations, nearest-neighbor geometric graphs, and aperiodic monotiles---demonstrate that each finite model converges to its corresponding PDE (posed on different manifolds like tori, disks, and spheres) in the limit of increasing vertex density.
\end{abstract}

\maketitle

This paper investigates the emergence of a macroscopic medium from a dense network of interconnected ``wires" or ``conduits." Discrete graphs with abstract relations between vertices have a long history with numerous applications \cite{Euler1736}. In contrast, we consider ``metric'' graphs with physical length along continuous edges that allows differential equations imposed over the entire system. Metric graphs appear in diverse settings, from optical fiber networks to neuronal structures, and in abstract contexts like quantum chemistry and dynamical systems.

\begin{figure*}
    \centering
    \includegraphics[width=0.85\textwidth]{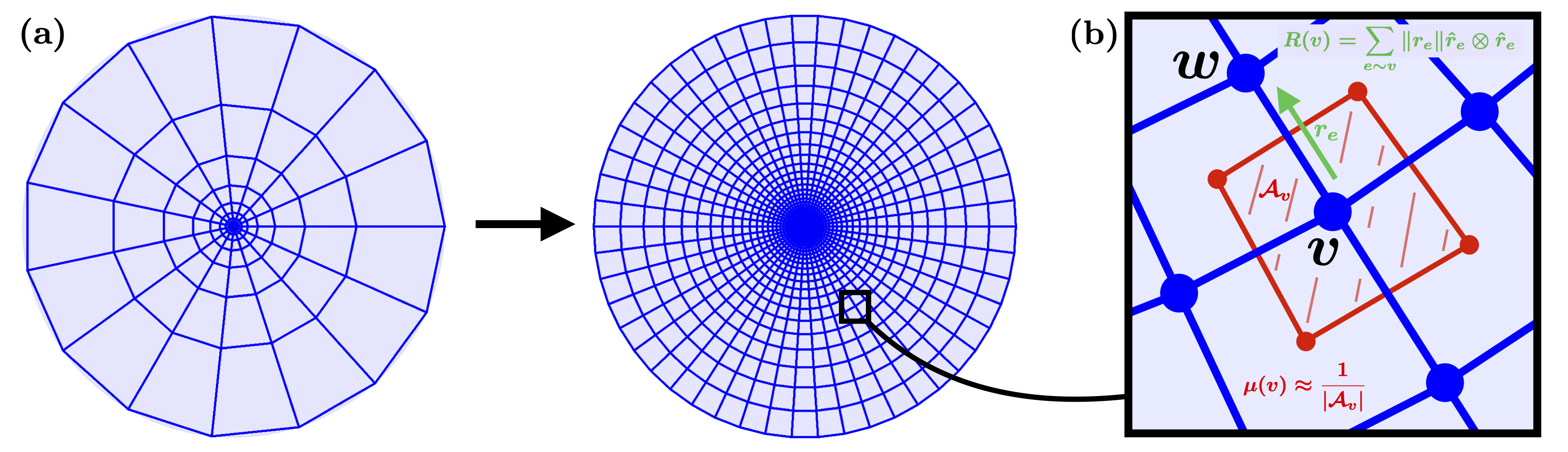}
    \caption{(a) The continuum-limiting procedure for the spiderweb within the unit disc. 
        (b) PDE quantities: the \emph{edge} tensor $R$ \eqref{eq:R}, and the \emph{vertex} density $\mu$ \eqref{eq:mu}.}
    \label{fig:density scheme} 
\end{figure*}

We define a spatial network \cite{Barthelemy2022} by embedding a metric graph in a manifold (e.g. Euclidean space, a torus or sphere). The dynamics on each edge obey one-dimensional differential equations with vertex boundary conditions that couple multiple edges. This results in a mesoscopic system that can be coarse-grained into a macroscopic PDE. The key question we address is:
\begin{quote}
\textit{What PDE arises from the continuum limit of a metric graph as vertices become dense and edge lengths shrink to zero? (Figure~\ref{fig:density scheme}a)}
\end{quote}

We answer this question for the weighted Laplace operator \eqref{eq:edge_de} via its corresponding generalized nonlinear eigensystem \cite{Colbrook2023}. The resulting PDE \eqref{eq:pde} leads to the more general equation \eqref{final bulk equation} after a discrete-to-continuum homogenization. The Laplace operator is a natural starting point due to its ubiquity in physics and engineering.  

There has been significant work studying differential equations on metric graphs, focusing mainly on quantum Hamiltonian Schr\"{o}dinger operators \cite{berkolaiko2013introduction}. Linus Pauling first considered differential equations on metric graphs to model free electrons in organic molecules \cite{pauling1936diamagnetic}. 
Recent examples include applications to nanomaterials \cite{Saremi2020, Lawrie2024} and wave propagation in microwave networks \cite{DoronE.1990Edoc}. Mathematical results cover nonlinear PDEs, spectral analysis, and thin-branched structures \cite{exner2008analysis, post2012spectral}. Applications also include plant and animal vasculature \cite{Katifori2010, Alim2013, Rocks2021, Qi2021} and cytoskeletal assemblies \cite{Yan2022}. More recently, neuroscience is a very promising area for graph models, with partial- and whole-brain connectomes becoming increasingly available \cite{Schlegel2024,Pospisil2024,MICrONSConsortium2025}.

However, little work addresses the continuum limit of high-density metric graphs embedded in manifolds. Previous studies have examined graphs with high symmetry \cite{cattaneo1997spectrum, exner2022continuum}, Ollivier-to-Ricci curvature convergence for edges with positive weights \cite{van2021ollivier}, and spectral dependence on shrinking edges \cite{Berkolaiko2024}, but a systematic treatment of general metric graph classes remains open.

\section*{Mathematical setup}

A \textit{metric graph} $G = (V,\,E)$ is a collection of vertices, $V$, and edges, $E$, equipped with local Euclidean coordinates $x\in[0_{e},\ell_{e}] \simeq e \in E$. We denote two adjacent vertices with the homogeneous relation $v \sim w$ (equivalently $w \sim v$) \cite{berkolaiko2013introduction}. We denote the corresponding edge $e_{vw}$ and drop the subscript when the context is evident. In the initial construction, we avoid loops and multiple edges between two given vertices. Later in our analysis, multiple edges enter the picture when ``periodizing'' graph partitions locally. Nevertheless, multiple edges are (in some ways) equivalent to a graph with reweighted edge lengths; see Theorem 1 in SM \cite{supp}. A real-valued scalar function, $f_{e}(x)$ for $x\in{e}$, is globally continuous if it is continuous on all individual edges and single-valued on all vertices. However, the derivatives of arbitrarily smooth functions along each edge generically disagree at the vertices. The Kirchhoff condition (below) quantifies the balance of multiple derivative values from incident edges. We derive a system of equations via an action principle, $\delta \mathcal{S} = 0$, where
\begin{align}
    \mathcal{S}(f) 
    \eeq
    \frac{1}{2}
    \sum_{e \in E} \int_{0_{e}}^{\ell_{e}} \! 
    \mathcal{L}_{e}(x,f_{e},f_{e}') \dd{x}, 
    \label{action}
\end{align} 
with edgewise Lagrangian function,
\begin{align}
\mathcal{L}_{e}(x,f_{e},f_{e}') \eeq  K_{e}(x)  \left| f_{e}'(x) \right|^{2} - \lambda\, C_{e}(x)\, |f_{e}(x)|^{2}.  
\label{Lagrangian}
\end{align} 
We initially assume general smooth edge ``conductivities'', $K_{e}(x)$ and ``capacities'', $C_{e}(x)$. The generality is useful in applications like biological transport networks \cite{Katifori2010}. 
However, the dense limit only requires leading-order constant mean values of $K_{e}$ and $C_{e}$ along each edge.
The stationarity condition $\delta \mathcal{S} = 0$ gives the edge equations, 
\begin{align}
    -\frac{1}{C_{e}(x)}\frac{d}{dx} \left( K_{e}(x)  f_{e}'(x)\right) 
    \eeq 
    \lambda \,  f_{e}(x).
    \label{eq:edge_de}
\end{align}
If the variation has support on a single vertex, then the \textit{Kirchhoff condition} at $x = v$ is
\begin{align}
    \sum_{e \sim v }  \hat{B}_{ve}\,  K_{e}(v) f_{e}'(v) 
    \eeq 
    0.
    \label{eq:kirchhoff_condition}
\end{align}
where the ``incidence matrix'' $\hat{B}_{ve} = \pm{1}$ gives the unit normal outward from edge $e$ at vertex $v$, and the sum is over edges $e$ connected to $v$ (denoted $e \sim v$).
The Kirchhoff condition, \eqref{eq:kirchhoff_condition}, shows why relevant functions usually have multi-valued derivatives at vertices.
The manifestly self-adjoint nature of the problem implies that the spectrum $\sigma(\lambda)$ is real-valued and discrete with $\lambda > 0$ \cite{berkolaiko2013introduction}. The system could represent (e.g.) the spatial component of diffusion ($\lambda \to -\partial_{t}$), wave dynamics ($\lambda \to -\partial_{t}^{2}$), or Schr\"{o}dinger evolution ($\lambda \to i\partial_{t}$).
At this point, we simplify to edgewise constant $C_{e}$, $K_{e}$, which applies in the eventual short-edge length limit. 
For a given edge $e$ with endpoints $v \simeq 0_{e}$, $w \simeq \ell_{e}$ and setting $k_{e} = \sqrt{\lambda\, C_{e} / K_{e}}$, the local solution of \eqref{eq:edge_de} is
\begin{align}
    f_e(x)
    \eeq 
    f(v)\, \frac{\Sin{k_{e} (\ell_{e}-x)}}{\Sin{k_{e} \ell_{e}}}
    +
    f(w)\, \frac{\Sin{k_{e} x}}{\Sin{k_{e}  \ell_{e}}},
    \label{eq:edge_eigenfunction}
\end{align}
which manifestly expresses continuity at the vertices $f_{e}(0_{e}) = f(v)$, $f_{e}(\ell_{e}) = f(w)$. 
Substituting into the vertex Kirchhoff condition \eqref{eq:kirchhoff_condition} at vertex $v$ gives
\begin{align}
    \sum_{e \sim v} k_{e} K_{e} \left( \Cot{k_{e} \ell_{e}} f(v) - \Csc{k_{e} \ell_{e}} f(w) \right) \eeq 
    0.
    \label{eq:kc}
\end{align}
This system over the set of vertices is a finite nonlinear eigenvalue problem (NEP) \cite{Gttel2017, Colbrook2023} with eigenvalues, $k$, and eigenvectors, $f(V)$ (see SM \cite{supp}).
Finally, \eqref{eq:kc} derives from the 1st-order stationarity of the finite quadratic action functional,
\begin{align}
    S(f) \eeq
    \frac{1}{2}
    \sum_{v \in V} \sum_{e \sim v} & k_{e} K_{e} \Big\{\! \Cot{k_{e}\ell_{e}} \left( f(v) - f(w) \right)^2 \nonumber \\
    & - 2\Tan{\tfrac{k_{e} \ell_{e}}{2}} f(v)f(w) \Big\},
    \label{eq:finite_functional}
\end{align}
which assumes $K_{e}$ and $C_{e}$ are constant along the edges. A similar functional still exists in the more general case, albeit in terms of generalized Green's functions in the place of $\Sin{k_{e} x}/\Sin{k_{e}\ell_{e}}$. 

\section*{Continuum limit}

We embed our metric-graph solution in a general Riemannian manifold, $M \supset G$.
The vertices are simply points, $v\in{M}$.
The parameterized edges of a metric graph embedded within a manifold need not be straight; a balled-up fishing net is the same as its laid-flat counterpart. 
In the continuum limit, both options will give the same result: for any two adjacent vertices, we take the limit as $\ell_{e}\to0$ with the vertices filling in densely within $M$. 
We assume $M$ has a Riemannian structure with a natural volume measure.
The graph requires a \textit{vertex number density}, $\mu(x)$, such that for all sufficiently coarse-grained subsets, $S\subseteq{G}$,
\begin{align}
    \int_{S} \mu(x) \dd{x} 
    \eeq \text{number of vertices} \, \in \, S.
\end{align}
We can use the empirical measure on the vertices, $\mu_{V}(x) = \sum_{v\in V}\delta(x-x_{v})$, which Portmanteau's theorem guarantees converges weakly to a continuum limit as the graph becomes dense. For an orthonormal basis of square-integrable functions, $\psi_{k}(x)\in\mathcal{L}^{2}(G)$,
\begin{align}
    \mu_{N}(x) \eeq \sum_{k\le N} \psi_{k}(x) \sum_{v \in V}  \psi_{k}(x_{v}) \, \approx \, \frac{1}{|\mathcal{A}_{v}|},
    \label{eq:mu}
\end{align}
for $x\in\mathcal{A}_{v}$, the dual cell area (volume in general) surrounding a given vertex (see figure~\ref{fig:density scheme}b).  
The spectral representation approximates the empirical measure for functions of bounded variation, which is also sufficient to define the weak derivative \cite{evans2018measure}. We derive a formal continuum limit $\ell_{e}\to0$ in \eqref{eq:finite_functional}. First, $\Cot{z} = z^{-1} + \mathcal{O}(z)$ and $2 \Tan{z/2} = z + \mathcal{O}(z^{3})$, as $z \to 0$. Then,
\begin{align}
    f(w)-f(v) \eeq r_{e}\cdot\grad{f(v)} + \mathcal{O}(\ell_{e}^{2}),
\end{align}
with $\hat{r}_{e} = r_{e}/\ell_{e}$ ($\ell_{e} = \|r_{e}\|$ by definition) the unit tangent pointing from $v$ to $w$ along their common edge, $e$. On a general manifold, $r_{e}$ is formally the tangent vector along a parameterized edge from $v$ to $w$. We impose a local Riemannian structure when stating $ \| \hat{r}_{e} \| = 1$ and supposing that $\ell_{e}$ is the geodesic distance along $e$. Finding a globally smooth extension of the metric for a general graph is beyond our current scope \cite{Nash1956}. We assume the deviation between edge curves and geodesics becomes negligible as $\ell_{e} \to 0$. Now comes the crux of the derivation. We define the following ``edge'' tensors at each vertex,
\begin{align}
    R_{K}(v) 
    & \eeq
    \sum_{e \sim v} \|r_{e} \| \, K_{e}\,\hat{r}_{e} \otimes \hat{r}_{e}, \\
    R_{C}(v) 
    & \eeq
    \sum_{e \sim v} \|r_{e} \| \, C_{e}\,\hat{r}_{e} \otimes \hat{r}_{e}.
\end{align}
Finally, to $\mathcal{O}(\ell_{e}^{2})$, \eqref{eq:finite_functional} becomes
\begin{align}
    S(f)
    \eeq 
    \frac{1}{2}
    \int_{G} & \left[ \grad f(x) \cdot R_{K}(x) \cdot \grad f(x) \right.
    \hspace{0.5cm} \notag \\ 
    & - \left. \lambda \, \tr{R_{C}(x)} f(x)^{2} \right] \mu(x) \dd{x}. 
\end{align}
Computing the variational derivative of the continuum functional produces the eigenvalue problem in the dense limit: 
\begin{align}
    -\frac{1}{\mu \, \tr{R_{C}}} \grad \cdot \left( \mu \, R_{K} \cdot \grad f \right)
    \eeq
    \lambda f, 
    \quad x \in G,
    \label{eq:pde}
\end{align}
where the associated ``natural" boundary condition is
\begin{align}
    \hat{n} \cdot  (\mu \, R_{K}) \cdot \grad f \eeq 0, 
    \quad x \in \partial G,
    \label{eq:free_condition}
\end{align}
where $\hat{n}$ is the unit normal to the boundary. Alternatively, $f|_{\partial G} = 0$ for pinned boundary vertices. 
For the examples, we simplify to the special case where $K_{e} = C_{e} = 1$ for all edges, $e$, and then $\lambda = k^2$. We define (see figure~\ref{fig:density scheme}b)
\begin{align}
    R(v)
    \eeq 
    \sum_{e \sim v} \| r_{e}\|\, \hat{r}_{e}\otimes \hat{r}_{e},
    \label{eq:R} \quad
    \tr{R(v)}
    \eeq 
    \sum_{e \sim v} \|r_{e}\|.
\end{align}

\section*{Riemannian Comparison}

\Eqref{eq:pde} gives genuinely distinct behavior from a traditional Laplace-Beltrami operator, with the main differences resulting from the trace, $\tr{R}$.
For example, for a completely homogeneous and isotropic graph, the only symmetric rank-2 tensor is proportional to the $d$-dimensional identity, $I_{d}$,
\begin{align}
    R 
    \eeq 
    \frac{\tr{R}}{d}\, I_{d} \quad \implies \quad   -\frac{1}{d}\,\Delta f
    \eeq 
    \lambda f.
    \label{eq:case0_pde}
\end{align}  
Uniform edge lengths imply $\tr{R} = \ell\,\deg(v)$ for every vertex. For this particular case, \eqref{eq:pde} reduces to the standard Laplace-Beltrami operator, \textit{but weighted by the spatial dimension of the graph embedding}.
Overall, the continuum ``graph material" can resemble many other materials but is (roughly speaking) less stiff by the dimension of the embedding space. Including globally constant conductivity and capacity, satisfying $K/C = d$, would cancel the anomalous diffusivity factor in \eqref{eq:case0_pde}.
On perfectly square lattices, previous work \cite{Exner2006, Pankrashkin2012, Nakamura2021, Exner2022} rigorously proves formal operator convergence to \eqref{eq:case0_pde}, commenting on the dimension factor. 
For the edge tensor, the summands are $\|r_{e}\|^{p} \, r_{e} \! \otimes r_{e}$ with $p=-1$.
The inverse Riemannian metric tensor roughly corresponds to $p = -2$.
The reduced system resembles the standard Laplace-Beltrami operator on a manifold with an augmented local metric tensor, $g(x)$.
However, there is no clear way (without adjusting $K_{e}$ and $C_{e}$) to put \eqref{eq:pde} into one-to-one correspondence with a Laplace-Beltrami operator, even if we impose conditions on the type of Riemannian metric. 
That is, the edge tensor is defined similarly to a discrete approximation of a Riemannian metric, but with a different distance scaling, \textit{i.e.}, with closest correspondence (see SM \cite{supp})
\begin{align}
    g(v)^{-1}
    \ \longleftrightarrow \ \frac{\text{dim}(M)}{\text{deg}(v)} 
    \sum_{e \sim v} \hat{r}_{e} \otimes \hat{r}_{e}.
\end{align}
However, for Riemannian metrics, $\tr{g} = \text{dim}(M)$ by definition. 

\section*{Axisymmetric spiderweb}

\begin{figure}[ht]
    \centering
    \includegraphics[width=\columnwidth]{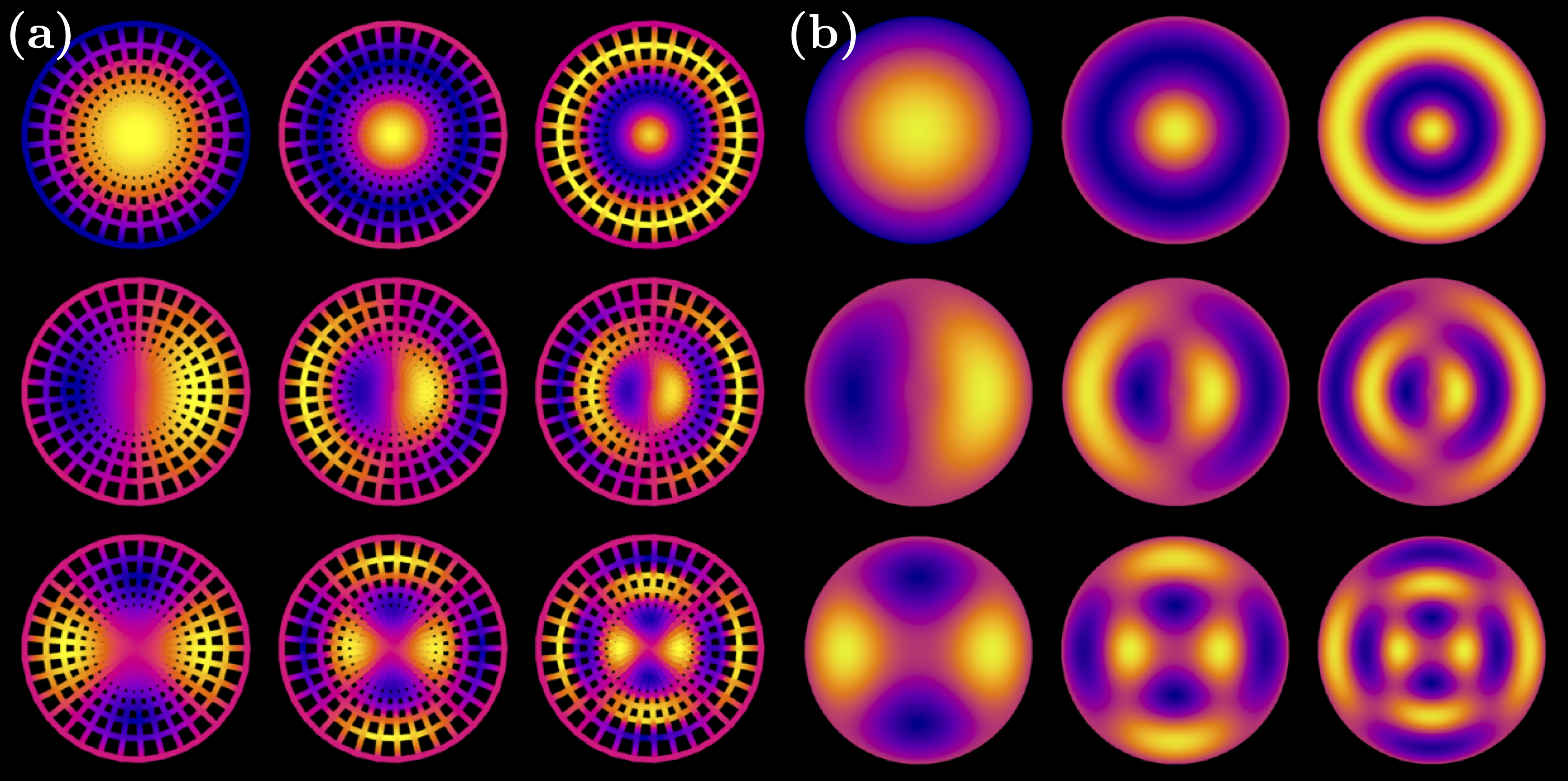}
    \caption{Continuum behavior for dense spiderwebs: 
    (a) \& (b) show graph eigenmodes \& PDE eigenmodes respectively for $(m,j)$ for $|V| \approx 500$; both cases only show the angular cosine (not sine) modes.} 
    \label{fig:dense spiderwebs}
\end{figure}

The first example, an axisymmetric ``spiderweb" (figure~\ref{fig:density scheme}a), shows the general validity of the continuum limit for an inhomogeneous and anisotropic graph. This simple example also illustrates the distinct nature of the graph limit compared to the standard Laplace-Beltrami operator. The spiderweb has small but finite local vertex spacings $dr$, $d\theta$ in each polar-coordinate direction $r$, $\theta$. Counting the central vertex at the origin, boundary vertices on the unit circle, and interior vertices gives a total of $|V| = (N_{r}-1)N_{\theta}+1$ where $N_{r}, N_{\theta}$ are the numbers of vertices in the radial and angular directions.
At interior vertices,
\begin{align}
    R(r) 
    & \eeq 
    R_{rr}\, \hat{r} \otimes \hat{r} + R_{\theta \theta}\, \hat{\theta} \otimes \hat{\theta}, 
    \\ 
    R_{rr} 
    & \eeq
    2 \, dr + 4\, r \sin^3( d\theta/2 ) = 2\, dr + \mathcal{O}(d\theta^{3}), 
    \\
    R_{\theta \theta} 
    & \eeq 
    \frac{r \sin^2( d\theta )}{ \Sin{ d\theta / 2}} = 2\,r \, d\theta + \mathcal{O}(d\theta^{3}).
\end{align}
The vertex density is $\mu = 1/\mathcal{A}_{v}$ with vertex-cell area $\mathcal{A}_{v} \approx r\,dr\,d\theta$.
The angular spacing is $d\theta = 2\pi/N_{\theta}$. 
The radial spacing depends smoothly on $r$ such that $dr = \rho(r)\,d\theta$. At the origin, $R(0) = \pi \rho(0)\, I$.
Angular symmetry allows Fourier solutions, $f(r,\theta) = f_{m}(r)e^{im\theta}+\text{c.c.}$ The continuum limiting PDE reduces to the ODE
\begin{align}
    r \rho(r) f_{m}''(r) - \left[ r (r + \rho(r)) k^{2} + m^{2} \right] f_{m}(r) 
    \eeq 
    0. 
    \label{eq:Fourier_spider}
\end{align}
We use pinned outer boundary conditions $f_{m}(r = 1) = 0$.
For center-point conditions, $f_{m}'(r = 0) = 0$ if $m = 0$ and $f_{m}(r = 0) = 0$ otherwise.
No choice of $\rho(r)$ renders the above operator into a Riemannian metric-based Laplacian. 
The original graph problem \eqref{eq:kc} also simplifies into decoupled radial systems (SM \cite{supp}).
The specific form $\rho(r) = r/\gamma$ allows analytical Bessel-like solutions. For $m\ne 0$, 
\begin{equation} 
    f_{j,m}(r) \, \propto \, \sqrt{r} J_{\nu}(\zeta_{j,\nu}\, r), \quad \nu = \sqrt{\gamma\,  m^2+\tfrac{1}{4}\, },
    \label{eq:Bessel_m_nonzero}
\end{equation}
where $J_{\nu}$ is the non-integer-order regular Bessel function. The spectral parameter, $k_{j,m} = \zeta_{j,\nu}/\sqrt{1+\gamma}$, with $J_{\nu}(\zeta_{j,\nu}) = 0$. The $m = 0$ solution uses the Neumann function,
\begin{equation}
    f_{j,0}(r) \, \propto \,  \Cos{  (j+\tfrac{1}{2})\pi \,r  }, \;  \; k_{j,0} 
    \eeq 
    \frac{(j+\tfrac{1}{2})\pi}{\sqrt{1+\gamma}}.
    \label{eq:Bessel_m_zero}
\end{equation}
We show numerical convergence of the graph solutions \eqref{eq:kc} to these continuum solutions in SM \cite{supp}.
Figures~\ref{fig:dense spiderwebs}(a)-(b) show visual comparisons between the graph and PDE modes.

\section*{Periodic examples}

This series of examples focuses on periodic graphs. We choose each periodic case study to highlight a particular phenomenon or subtlety. We build up a surplus of technical tools required to handle more general graphs in random cases. Homogenization for random graphs is based heavily on periodizing local partitions into local lattice structures. 

\begin{figure*}
    \centering
    \includegraphics[width = \textwidth]{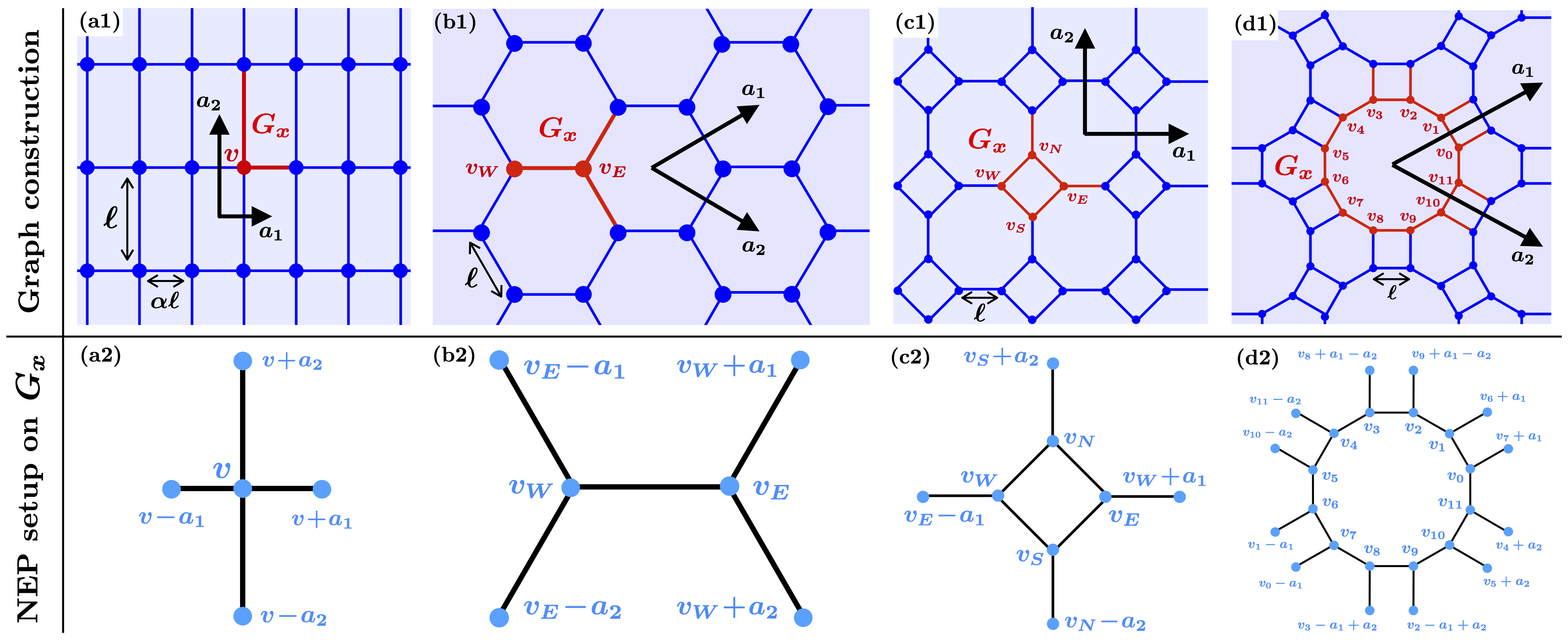}
    \caption{(a1)-(d1) Illustrations of periodic graph constructions. 
    The graph vectors $a_{1}, a_{2}$ translate the unit graphs $G_{x}$ (in red). 
    (a2)-(d2) The graph NEP is fully specified by the vertices and neighbors of $G_{x}$.
    Construction and NEP setup for the truncated square graph. (a) The orthogonal length $(1 + \sqrt{2})\ell$ lattice vectors $a_{1}$ and $a_{2}$ translate the diamond-shaped unit graph $G_{x}$. (b) The base vertices in $V_{x} = \{v_{N}, v_{S}, v_{E}, v_{W}\}$ each have three neighbors: two base vertices and one translated base vertex.} 
    \label{fig:constant_periodic_graphs}
\end{figure*}

\subsection*{Lattice Graph Construction}\label{sec: lattice}

We construct a $d$-dimensional periodic graph by translating copies of a ``unit" graph $G_{x}$ with ``base" vertices $v \in V_{x}$. Given a fundamental basis spanning $\mathbb{R}^{d}$, $\{a_{1}, \ldots, a_{d}\}$,  we define a lattice as the set, 
\begin{align}
    \Lambda
    \eeq 
    \left\{ \, \sum_{i = 1}^{d} n_{i} \, a_{i} \ ; \ n_{i} \in  \mathbb{Z} \right\}.
\end{align} 
The full periodic graph is $G = G_{x} + \Lambda$. 
For example, the 2D square graph has unit graph $G_{x}$ comprising a single vertex in $V_{x} = \{v\}$, and two orthogonal edges defined by $a_{1} = (\ell,0)$, $a_{2} = (0,\ell)$ (see figure~\ref{fig:constant_periodic_graphs}(a1)).
For periodic graphs, solving the full-graph NEP requires information only from $G_{x}$. The vertex condition (\eqref{eq:kc}) at $v \in V_{x}$ with neighbors $w \in V_{x}$ is
\begin{align}
    \sum_{r \in \Lambda} \sum_{w + r \sim v }\big[ & \Cot{k\, \ell_{v,w + r } } \, f(v) - \nonumber \\ & \Csc{k\, \ell_{v,w+ r } } \, f(w+ r ) \big]
    \eeq 
    0,
    \label{eq:periodic_NEP}
\end{align}
where $\ell_{v,w+r}$ is the length of the edge connecting $v$ and $w+r$.
We assume plane-wave solutions $f(x) = \hat{f}(\kappa)\, e^{i \kappa \cdot x} + \text{c.c.}$.
The wave vectors, $\kappa$, lie in the dual lattice, $\Lambda^{\!*}$, with basis elements, $a^{*}$, satisfying $a^{*}_{i} \cdot a_{j} = \delta_{i,j}$. We adopt the short-hand, $\vartheta_{i} = \kappa\cdot a_{i}$, for the phase angle throughout the following sections.
We also define $\kappa_{i} = \vartheta_{i} / \ell = \mathcal{O}(1)$ as $\ell \to 0$.
The embedding of $\Lambda$ into the $d$-dimensional torus, $\mathbb{T}^{d}$, happens by taking $\kappa$ to satisfy various quantization conditions. The solutions are periodic 
$f(x + r_{i} ) = f(x)$ with respect to a set of $d$ linearly independent lattice vectors $r_{i} \in \Lambda$. Therefore, $\kappa \cdot r_{i} = 2\pi n_{i}$ for $n_{i} \in \mathbb{Z}$.

\subsection*{Rectangular and Square Lattices}

Figure~\ref{fig:constant_periodic_graphs}(a1) shows a periodic rectangular graph. The fundamental lattice vectors are $a_{1} = (\alpha \, \ell, 0)$ and $a_{2} = (0, \ell)$ with uniform edge length $\ell > 0$ and anisotropy parameter $\alpha > 0$. Figure~\ref{fig:constant_periodic_graphs}(a2) shows $v$ has four neighbors: $\{v \pm a_{1}, v \pm a_{2}\}$. In terms of Fourier parameters, \eqref{eq:periodic_NEP} becomes 
\begin{align}
    \Cot{k \ell_{1}} + \Cot{k \ell_{2}} 
    \eeq & 
     \frac{\Cos{\vartheta_{1}}}{\Sin{k \ell_{1}}} + \frac{\Cos{\vartheta_{2}}}{\Sin{k \ell_{2} }}.
     \label{eq:rectangle_dispersion_relation}
\end{align} 
For $\alpha = 1$, the dispersion relation simplifies to 
\begin{align}
\Cos{k \ell}
\eeq 
\frac{\Cos{\vartheta_{1}} + \Cos{\vartheta_{2}}}{2}. \label{eq:square_dispersion_relation}
\end{align} 
\Eqref{eq:square_dispersion_relation} is particularly intriguing: it is the Pythagorean Theorem for a right triangle on the \textit{surface of a sphere} with spherical side angles $|\vartheta_{1} \pm \vartheta_{2}|/2$ and hypotenuse angle $k\ell$. Nontrivial geometric structure in spectral space can lead to interesting topological phenomena \cite{berry1984quantal}.
For \eqref{eq:rectangle_dispersion_relation} as  $\ell \to 0$, 
\begin{align}
k^{2} \eeq \frac{\kappa \cdot R \cdot \kappa}{\tr{R}}.
\end{align}
The density is constant, $\mu(x) = \alpha^{-1} \ell^{-2}$, and 
\begin{align}
    R(v)
    & \eeq
    2 \left(a_{1} \otimes a_{1} + a_{2} \otimes a_{2}\right)
    \eeq
    2\ell
    \begin{bmatrix}
        \alpha & 0 \\
        0 & 1
    \end{bmatrix}.
\end{align}
The corresponding PDE agrees with the limiting NEP,
\begin{equation}
    - \left(\frac{\alpha}{\alpha + 1} \partial_{x}^{2} + \frac{1}{\alpha + 1} \,\partial_{y}^{2} \right)f
    \eeq 
    \lambda f,
\end{equation}
which matches \cite{Tzella2016}, who model advection-diffusion on a rectangular grid.
By approximating the concentration in the long-time limit with a Gaussian diffusion kernel, they find an effective zero-advection diffusivity tensor, $D$, near the source (\cite{Tzella2016} eqs.~(12-14)) 
\begin{align}
    D 
    \eeq 
    \frac{R}{\tr{R}}
    \eeq 
    \frac{1}{\alpha + 1} 
    \begin{bmatrix}
        \alpha & 0 \\ 
        0 & 1
    \end{bmatrix}.
\end{align}
Furthermore, they find a medium-range diffusion kernel based on the squared Manhattan distance, $\|x\|_{1}$, rather than the rotationally invariant Euclidean metric, $\|x\|_{2}$. However, the general bound $\|x\|_{1}^{2}\le{d}\,\|x\|_{2}^{2}$ implies the results agree in the full continuum limit.

\subsection*{Hexagonal Lattice}

Figure~\ref{fig:constant_periodic_graphs}(b1) shows the first case with more than one distinct base vertex, $V_{x} = \{v_{E}, v_{W}\}$, with displacement vector $r_{WE} = (\ell, 0)$. Figure~\ref{fig:constant_periodic_graphs}(b2) shows the adjacency structure of $G_{x}$.
The fundamental lattice vectors are $a_{1}, \, a_{2} = \frac{\ell}{2} \left( 3, \, \pm\sqrt{3} \right)$.
In this case, \eqref{eq:periodic_NEP} is a $2 \times 2$ system over $V_{x}$,
\begin{align}
    \begin{bmatrix}
        -3\gamma & 1 + e^{i \vartheta_{1}}  + e^{i \vartheta_{2}} \\ 
        1 + e^{-i \vartheta_{1}}  + e^{-i \vartheta_{2}} & -3\gamma
    \end{bmatrix} \! \! 
    \begin{bmatrix}
        f_{E} \\ 
        f_{W}
    \end{bmatrix}
    \eeq 
    0.
    \label{eq:hexagonal_NEP_system}
\end{align}
The dispersion relation is,
\begin{align}
    \Cos{k \ell} 
    \eeq  \pm  \frac{\left|1 + e^{i \vartheta_{1}}  + e^{i \vartheta_{2}} \right|}{3}. \label{gammaH}
\end{align}
In this case, $\ell \to 0$ stills leads to, 
\begin{align}
    k^{2} 
    \eeq 
    \frac{\kappa\cdot \kappa}{2}
    \eeq  \frac{\kappa_{1}^{2}+\left(\kappa_{1}-\kappa_{2}\right)^{2}+\kappa_{2}^{2}}{9} \label{hex k kappa leading order}.
\end{align}
For the direct continuum model, the neighbor vectors,   $r_{j} = \ell (\Cos{j\pi/3}, \Sin{j \pi/3})$. A ``West'' vertex, $v_{W}$, links to three ``East'' vertices at $v_{W} + r_{j}$ with $j \in \{0, 2, 4\}$. Likewise, an East vertex, $v_{E}$, links to three West vertices at $v_{E} + r_{j}$ for $j \in \{1, 3, 5\}$.
In both cases, we have $R(v_{W}) = R(v_{E}) = 3/2 \ell I$.
Because $\mu(v) = 2\ell^{-2}/\sqrt{27}$ and $R_{E} = R_{W} \propto I$, we again have \eqref{eq:case0_pde}.

\subsection*{Truncated Square and Trihexagonal Lattices}\label{subsubsec: Truncated square graph}

The hexagonal graph has a constant $R$ tensor despite having more than one type of vertex neighborhood. 
Here, we present a slightly more complicated example with a well-defined continuum limit but without a well-behaved $R$ tensor.
Figure~\ref{fig:constant_periodic_graphs}(c1) shows a truncated square graph with $V_{x} = \{v_N, v_S, v_E, v_W\}$. In this case, \eqref{eq:periodic_NEP} is a $4 \times 4$ system over $V_{x}$,   
\begin{align}
    \begin{bmatrix}
        -3\gamma & e^{i\vartheta_{2}}  & 1 & 1 \\ 
        e^{-i\vartheta_{2}} & -3\gamma & 1 & 1 \\
        1 & 1 & -3\gamma & e^{i\vartheta_{1}} \\ 
        1 & 1 & e^{-i\vartheta_{1}} & -3\gamma
    \end{bmatrix}\!\!
    \begin{bmatrix}
        f_{N} \\ 
        f_{S} \\ 
        f_{E} \\ 
        f_{W} 
    \end{bmatrix}
    \eeq 
    0,
    \label{NSEW system}
\end{align}
with lattice basis vectors $a_{1} \eeq (\alpha \, \ell , 0)$ and $a_{2} \eeq (0, \alpha \, \ell)$ where $\ell$ is the edge length and $\alpha = 1 + \sqrt{2}$. The dispersion relation is, 
\begin{align}
81 \gamma^{4}-54 \gamma^{2}-24\, \gamma_{\text{C}}\, \gamma -4 \gamma_{\text{O}}+1 = 0.
\end{align}
where $\gamma = \Cos{k \ell}$ and we define 
\begin{align}
\gamma_{\text{C}} = \frac{\cos(\vartheta_{1}) + \cos(\vartheta_{2})}{2} , \quad  
\gamma_{\text{O}} = \cos(\vartheta_{1}) \cos(\vartheta_{2}).
\end{align}
As expected, the truncated square lattice reduces to the cardinal square lattice as the diagonal edges shrink to zero, and all the NSEW vertices become a single vertex. Likewise, the lattice reduces to the ordinal case as the vertical and horizontal edges shrink to zero. 
This particular lattice is the first example of behavior that does not fit directly within the homogeneous $R$-tensor framework. In particular, as $\ell \to 0$, 
\begin{align}
    k^{2}
    \eeq 
    \frac{(1+\sqrt{2})^{2}}{6}  \frac{\kappa\cdot \kappa}{2} \, \approx \, 0.97 \, \frac{\kappa\cdot \kappa}{2}.
    \label{truncated square k-squared limit}
\end{align}
The limiting dispersion relation suggests a different scaling of the standard Laplace operator than anything up to this point.
To obtain a simple rescaled Laplace operator, we require $R \propto I$, which gives $d^{-1} \Delta$ after dividing out by the trace.
However, the $R$ tensor has no apparent well-defined pointwise limit in this case.
For the edge vectors, $r_{j} = (\Cos{j\pi / 4}, \Sin{j\pi / 4})$, North vertices have $j \in \{2,5,7\}$; South has $j \in \{1,3,6\}$; East has $j \in \{0,3,5\}$; and West has $j \in \{1,4,7\}$.
Therefore,
\begin{align}
    R(v_{N}) \eeq 
    R(v_{S}) 
    \eeq 
    \ell
    \begin{bmatrix}
        1 & 0 \\
        0 & 2
    \end{bmatrix}, \\
    R(v_{E}) \eeq 
    R(v_{W})
    \eeq 
    \ell
    \begin{bmatrix}
        2 & 0 \\
        0 & 1
    \end{bmatrix}.
\end{align}
These represent an $\bigo{\ell}$ oscillation in $R$ at the scale of a vertex neighborhood, with no obvious well-defined limit as $\ell \to 0$. However, the overall system has a well-defined continuum behavior seen in \eqref{truncated square k-squared limit}. For a more complicated example, the truncated trihexagonal graph in figure~\ref{fig:constant_periodic_graphs}(d1) (also see SM \cite{supp}) reduces to 
\begin{align}
    k^{2} \, = \, \frac{(3+\sqrt{3})^{2}} {24} \frac{\kappa\cdot \kappa}{2} \, \approx \, 0.93 \, \frac{\kappa\cdot \kappa}{2}.
\end{align}
In the following section, we resolve the issue using a mixed discrete-continuum \emph{homogenization} technique to generate a well-defined renormalized local $R$ tensor.

\section*{Homogenization}

We express $f(v)$ as an unknown function $F(x)$ of the macroscale, $x$ (defined below), displaced by an also unknown microscale ``corrector'' vector $\xi_{v}$ \cite{Berlyand2018}, 
\begin{align}
    f(v) \eeq  F( x )  + \xi_{v} \cdot \grad_{\! x}F(x),
\end{align} 
where $\grad_{\! x}$ is the macroscale gradient. To first order,
\begin{align}
    f(w) - f(v) 
    \eeq 
    \left(r_{xy} + \xi_{vw} \right) \cdot \grad_{\! x}F(x) + \mathcal{O}(r^{2}, \xi^{2}),
\end{align}
where, $v\in x$, $w \in y$, $r_{xy}$ is the tangent vector pointing from $x$ to $y$, and $\xi_{vw} = \xi_{w} - \xi_{v}$.

\subsection*{Graph Covering and Periodization}

\begin{figure*}[t]
    \centering
    \includegraphics[width=\textwidth]{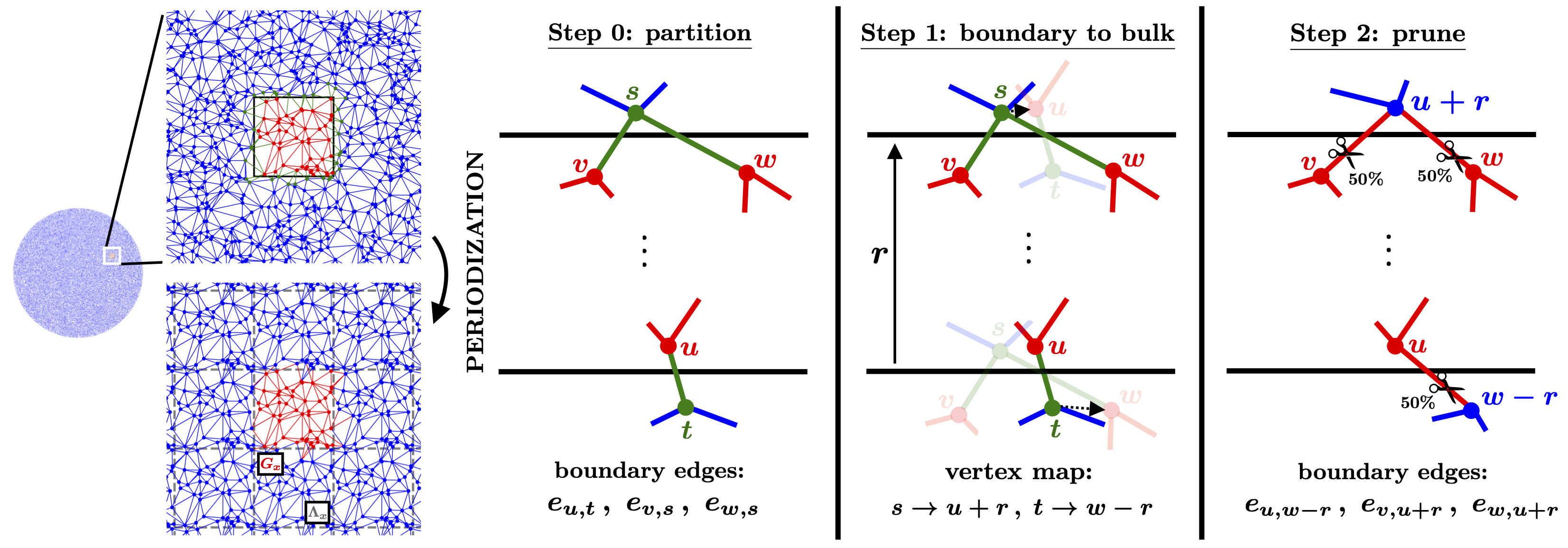}
    \caption{Illustration of the patching and periodization procedures: (a) For a graph $G$ within an embedding space $M$, we extract a subgraph patch (red) and boundary vertices and edges (green). 
    We periodize the subgraph within the patch, producing the top-right periodic graph. The periodic graph comprises a base graph, $G_{x}$, translated by the lattice, $\Lambda_{x}$, analogous to (e.g.) \figref{fig:constant_periodic_graphs}(a). In the top-right, we show $3 \times 3$ repeated copies of $G_{x}$ (blue). 
    (b) After partitioning into patches, the periodization procedure identifies boundary vertices with bulk (internal) vertices. Each boundary vertex (green) maps to the closest bulk vertex (red) after wrapping around the patch, with the mapping in \eqref{w to x connection}. 
    Penultimately, the periodic edges have twice the density of the bulk edges. We, therefore, prune each periodic edge with 50\% probability.}
    \label{fig:patching}
\end{figure*}

To specify macro and micro scales, we cover the vertex set, $V$, with a collection of connected subsets,
\begin{align}
\mathcal{X} = \{ x \subseteq V \}, \quad \text{with} \quad V = \bigcup_{x \in \mathcal{X}} x.
\end{align}
Technically, $\mathcal{X}$ is a discrete \textit{coordinate atlas} where we blur the distinction between covering subsets and coordinate maps; \textit{i.e.}, we refer to the covering subsets by coordinates that we later use in the continuum limit. The patches can either overlap or non-intersect. In parctice, we ensure all subsets contain the same number of vertices. Each patch comprises a volume of the embedding space, giving a local estimate for the vertex number density, $\mu(x)$.
With these definitions, the double summation in the bilinear form becomes 
\begin{align} 
    \label{eq:sum_partitioning}
    \sum_{v \in V}\sum_{w \sim v}
    \eeq
    \sum_{x,y \in \mathcal{X}} \sum_{ v \in x } \sum_{ (w  \in y) \sim v } \rho_{v} \, \rho_{w}
\end{align}
with the discrete ``partition of unity'' $\rho_{v} = 1/|\mathcal{X}_{v}|$ where $\mathcal{X}_{v} = \{x \in \mathcal{X} \,:\, v \in x\}$ (see SM \cite{supp}). 
The homogenization procedure requires independent decoupled sums at each local $x \in \mathcal{X}$ with suitable boundary conditions for edges that cross over multiple patches. Periodicity gives the most reliable results in this regard \cite{Davies2002}. 
There is no unique way to \textit{``periodize''} a given patch, $x \in \mathcal{X}$. 
We outline a procedure as follows (illustrated in figure~\ref{fig:patching}), using the unit graph and lattice framework defined for the periodic graphs in figure~\ref{fig:constant_periodic_graphs}.
We assume the unit graph to comprise the vertices and edges involving each $v \in x$ (in red in figure~\ref{fig:patching}), as well as the boundary vertices $(w \in y) \sim x$ and corresponding edges (in green in figure~\ref{fig:patching}). 
We impose a local lattice structure, $\Lambda_{x}$, where $r_{xy} \in \Lambda_{x}$. 
For the periodic graphs, the connection of the unit graphs $G_{x}$ to their $\Lambda$-translated copies was self-evident.
For more general graphs, we identify each $w \in y$ with its nearest neighbor after periodically wrapping it around the fundamental base cell. That is, for each $(w \in y) \sim x$, we make the identification
\begin{align}
    w \, \to\,  \text{argmin}\, \| x -  (w  \text{ mod } \Lambda_{x}) \|. \label{w to x connection}
\end{align}
The vertex $v \in x$ which was previously connected to $w \in y$ is now connected to another vertex in $x$ translated by $\Lambda_{x}$.
The last complication is that the number of boundary edges is, statistically, twice as large as the original boundary distribution. To remedy the double counting issue, we remove every boundary edge with probability $1/2$. For periodic graphs, we only remove doubled vertices. 
We finally have a periodic graph $G_{x} + \Lambda_{x}$ derived from the local unit graph $G_{x}$.
On a curved manifold (e.g. a sphere), the $r_{xy} \in \Lambda_{x}$ condition requires that the lattice varies from patch to patch.
The boundary-edge reattachment and pruning procedure advantageously removes large-scale imbalances in the number of edges. Large-scale linear gradients are the biggest distinction between periodic and nonperiodic systems. The periodization procedure works well if there is a bound on the degree of large-scale imbalance in the edge statistics. 

\subsection*{Discrete Homogenization}

With the periodization, the exact sum correspondence in \eqref{eq:sum_partitioning} maps to a sum over locally periodic patches. The periodic copies produce a system with $\rho_{v} = 1$ by definition. That is, the vertices in neighboring patches, $(w \in y) \sim x$, are replaced with images $w + r \in x$ for $r \in \Lambda_{x}$.
Moreover, the tangent vectors pointing from $x$ to $y$ also map to lattice vectors, $r_{xy} \in \Lambda_{x}$. We define the conductivity- and capacity-weighted edge lengths on each edge, $e$, as $\ell^{(K)}_{e} = \frac{\ell_{e}}{K_{e}}$ and $\ell^{(C)}_{e} = C_{e}\,\ell_{e}$.
Altogether, the two-scale bilinear form becomes  
\begin{align}
    S(F,\xi) = \sum_{ x \in \mathcal{X}} \! \! \left(\, \| \grad_{\! x}F(x) \|_{\xi,K}^{2} - \lambda\,  \mathcal{C}(x) \, F(x)^{2} \,
    \right),
\end{align}
where,
\begin{align}
\| \grad_{\! x}F \|_{\xi,K}^{2} = \frac{1}{2}
\sum_{r\in \Lambda_{x} } \sum_{ v \in x } \sum_{ w + r \sim v } \! \! \frac{|(r+ \xi_{vw}) \cdot \grad_{\! x}F |^{2}}{\ell^{(K)}_{v,w+r}},
\label{eq:Sx(DF,xi)}
\end{align}
and 
\begin{align}
    \mathcal{C}(x) 
    \eeq
    \frac{1}{2}
    \sum_{ r\in \Lambda_{x} } \sum_{ v \in x } \sum_{ w +r \sim v }\ell^{(C)}_{v,w+r} \label{Tx sum}.
\end{align}
Extremizing \eqref{eq:Sx(DF,xi)} with respect to $\xi_{v}$ for arbitrary $\grad_{\! x}F$, we obtain 
\begin{align}
    \sum_{ r \in \Lambda_{x} } \sum_{ w + r \sim v } \frac{ r+ \xi_{w} - \xi_{v} }{\ell^{(K)}_{v,w+r}} \eeq
    0. \label{eq:metric_graph_cell_problem}
\end{align}  
\Eqref{eq:metric_graph_cell_problem} is a discrete analogue of asymptotic homogenization for PDEs \cite{Allaire1992, Nguetseng2003, Nguetseng2004}. In particular, \Eqref{eq:metric_graph_cell_problem} is called the ``cell problem'' or ``corrector equation'' for regularizing continuous elliptic operators with rapidly varying non-constant coefficients. 
We derive a compact form of \eqref{eq:metric_graph_cell_problem} using Hilbert spaces over the vertex and edge states within $x \in \mathcal{X}$ in SM \cite{supp}. 
Recent work also exists on discrete versions of the homogenization problem \cite{Biskup2011, Gloria2011}. To our knowledge, no work on homogenization yet exists on the continuous limit for metric graphs. 
With all definitions in place for $x \in \mathcal{X}$, we define the regularized action functional 
\begin{align}
    \mathcal{S}(F) \eeq \frac{1}{2}\int_{M} \big[ \grad_{\! x}F(x) \cdot \mathcal{K}(x) \cdot \grad_{\! x}F(x) - 
    \nonumber \\ 
    \lambda\, \mathcal{C}(x)\, F(x)^{2} \big]  \, \mu(x) \dd{x} \label{limiting action},
\end{align}
where 
\begin{align}
    \mathcal{K}(x)
    \eeq 
    \sum_{r\in \Lambda_{x} } \sum_{ v \in x } \sum_{ w + r \sim v } \frac{(r + \xi_{vw}) \otimes (r + \xi_{vw})}{\ell_{vw}^{(K)}}.
\end{align}
Further, $\mu(x) = |x|/\mathrm{vol}(x)$ is the local vertex number density where $|x|$ is the number of vertices in $x$ and $\mathrm{vol}(x)$ is the volume that $x$ occupies within the manifold, $M$. Taking the variational derivative once more,
\begin{align}
    - \frac{1}{\mu\, \mathcal{C}}\grad_{x} \cdot (\mu \, \mathcal{K} \cdot \grad_{x} F) = \lambda\, F. \label{final bulk equation}
\end{align}
Finally, as a  measure of the homogenization we define the scalar \textit{diffusivity} and \textit{effective dimension},
\begin{align}
    \Diffu \eeq \frac{\tr{\mathcal{K}}}{\mathcal{C} d}, \qquad  d_{\text{eff.}} \eeq \frac{1}{\Diffu}  \label{eq:homogenized_tensor_coefficient} 
\end{align}
In the perfectly homogenous examples that do not require homogenization, $d_{\text{eff.}} = d$, where $d$ is the spatial dimension. Otherwise, $d_{\text{eff.}}$ gives an analogous fractal dimension for the underlying graph.
In SM \cite{supp}, we rederive the anomalous diffusivity from \eqref{truncated square k-squared limit} for the truncated square graph.

\section*{Random graphs} 

This section derives continuum limits for random graph structures that approximate real-world networks.
These examples require homogenization.
We investigate three graph structures: uniformly random Delaunay triangulations \cite{Lee1980}, uniformly random geometric graphs (each vertex connects to all others within a given neighborhood) \cite{Penrose2003}, and aperiodic monotiles \cite{Smith2024}, as illustrated in figures~\ref{fig:graphs_and_coefficients}(a1), (b1) and (c1) respectively.
By solving ensembles of cell problems (\eqref{eq:metric_graph_cell_problem}) for increasing vertex densities, we estimate the limiting forms of the homogenized quantities, $\mathcal{K}$ and $\mathcal{C}$, and construct the limiting PDE (from \eqref{limiting action}).  
Through a series of specific graph NEPs, we show the convergence of the graph eigenmodes to the PDE eigenmodes.

\subsection*{Local Homogenization}\label{sec: Continuum limits}

\paragraph{Uniformly random Delaunay triangulation} 
We first uniformly randomly sample $\left[-1, 2\right]^{2} \subset \mathbb{R}^{2}$ until $|V_{n}|$ vertices lie within the unit square $\left[0, 1\right]^{2}$.
We Delaunay triangulate the full set of vertices and then perform the patching and periodization procedure (illustrated in figure~\ref{fig:patching}) on the unit square to construct a cell problem (\eqref{eq:metric_graph_cell_problem}) of size $|V_{n}| \times |V_{n}|$. 
\begin{figure}[t]
    \centering
    \includegraphics[width = \columnwidth]{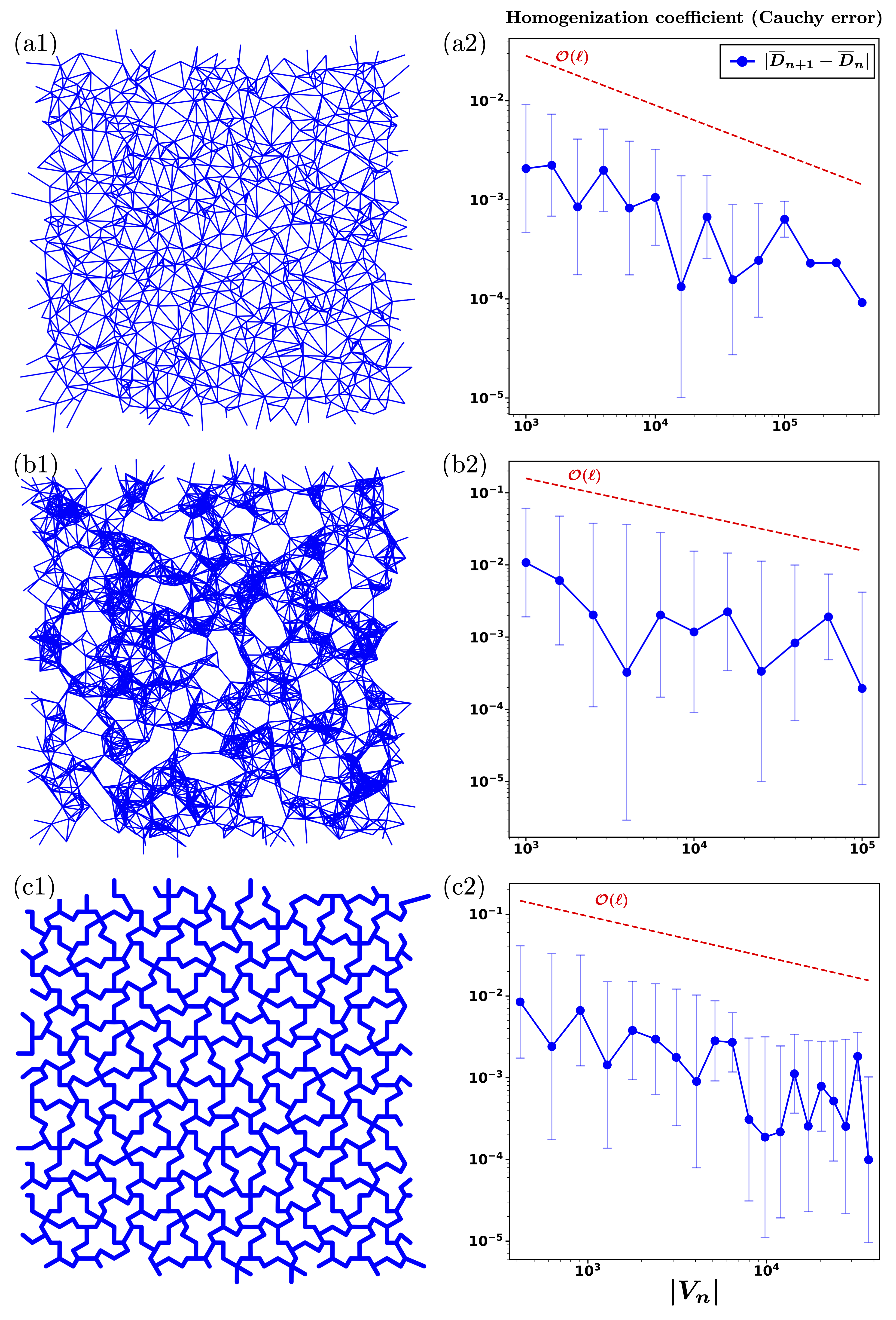}
    \caption{Figures (a1), (b1), and (c1) (left-side) illustrate the three different graph types: random Delaunay triangulations, random geometric graphs, and aperiodic monotiles. Each left-side image represents a periodized patch (described in figure~\ref{fig:patching}). 
    Figures (a2), (b2), and (c2) (right-side) show Cauchy convergence of the homogenization diffusivity coefficient, $\Diffu$, \eqref{eq:homogenized_tensor_coefficient}. We plot $|\text{mean}(\Diffu_{n + 1}) - \text{mean}(\Diffu_{n})|$ for $|V_{n + 1}|/|V_{n}| = 10^{1/5}$. We push the Delaunay triangulation to $|V_{n}| = 10^{6}$, but without multiple realizations beyond $10^{5}$.
    We display log-space uncertainties as defined in SM \cite{supp}.}
    \label{fig:graphs_and_coefficients}
\end{figure}
With the correctors $\xi_{v}$, we construct $\mathcal{K}_{n}$ and $\mathcal{C}_{n}$.
We define the scaled trace-free \textit{deviatoric} conductivity (with $d = 2$),
\begin{align}
    \Sigma \eeq \frac{\, \mathcal{K}}{\tr{\mathcal{K}}}  - \frac{1}{d}I, \quad \| \Sigma \|^{2}_{2} \eeq \frac{\tr{\mathcal{K}^{2}}}{\tr{\mathcal{K}}^{2}} - \frac{1}{d}. \label{deviatoric}
\end{align}
Empirically, $\| \Sigma_{n} \|^{2}_{2} = \bigo{|V_{n}|^{-1}}$ for increasing $n$ (see SM \cite{supp}).
We measure the mean homogenization coefficient $\overline{\Diffu}_{n}$ over ensembles of random realizations of patches of size $|V_{n}|$.
Figure~\ref{fig:graphs_and_coefficients}(a2) shows the Cauchy-error, $\mathcal{E}_{n} = |\overline{\Diffu}_{n + 1} - \overline{\Diffu}_{n}|$,  versus $|V_{n}|$, where $|V_{n+1}|/|V_{n}| = 10^{1/5}$ and $|V_{1}| = 10^{3}$.
We find approximately $\mathcal{E}_{n} = \bigo{|V_{n}|^{-1/2}}$, where the edge lengths are roughly the inverse square root of the vertex density, $\ell_{n} = \bigo{|V_{n}|^{-1/2}}$. 
The final mean effective coefficient goes to 11, where $V_{11} = 10^{5}$ and $\overline{\Diffu}_{11} \approx 0.9201 / 2$.
Altogether, the continuum limiting PDE takes the form
\begin{align}
    - \Diffu \, \Delta f
    \eeq 
    \lambda f,
    \label{eq:random_continuum_limit} 
\end{align}
where we approximate the theoretical limit $\Diffu \approx \Diffu_{11}$. In this case, we find an effective dimension, $d_{\text{eff.}} \approx 2.1737$.

\paragraph{Uniformly random geometric graph} We use the randomly sampled points from the Delaunay triangulation and instead connect each vertex to every neighbor within distance $2 \ell_{n}$.
We reuse the same graph sizes from the Delaunay triangulation. Figure~\ref{fig:graphs_and_coefficients}(b2) shows the Cauchy convergence of $\Diffu_{n}$, with $\overline{\Diffu}_{11} \approx 0.6781/2$ and an effective dimension, $d_{\text{eff.}} \approx 2.9494$.

\paragraph{Aperiodic monotile}  
We construct a large aperiodic monotile (\textit{i.e.}, $|V| \approx 10^{5}$, which is computationally intensive) and periodize patches (subgraphs) of increasing area.
We obtain ensembles of graphs of approximately equal vertex sizes by taking different patch locations within the larger graph.
We choose $|V_{1}| \approx 400$ (with small fluctuations due to graph sizes being chosen by area), $|V_{13}| \approx 15\,000$ and $|V_{n+1}| / |V_{n}| \approx 10^{1/10}$.
Figure~\ref{fig:graphs_and_coefficients}(c2) shows the Cauchy-error convergence with $\overline{\Diffu}_{13} \approx 0.6074/2$ and an effective dimension, $d_{\text{eff.}} \approx 3.2927$.
\paragraph{Effective dimension} The effective dimension is an intriguing diagnostic to compare different graph structures. For the original lattices, we find $d_{\text{eff.}}$ is always only slightly larger than the true dimension. The same is also the case in the Delaunay triangulation. The Delaunay triangulation is valuable for its well-behaved properties, and we seem to find that here. The effective dimension is roughly similar to the regular (albeit somewhat stranger) lattices. 
On the other hand, we see that the geometric graph and (especially) the monotile are much more consistent with $d_{\text{eff.}} \approx d+1$. We speculate that the effective dimension hints at a more well-behaved embedding in three (or even higher) dimensions, for which the current graphs are two-dimensional projections. If so, then the two-dimensional modes considered here would only be those invariant in the higher dimensions, but non-invariant large-scale modes would also exist. 

\subsection*{Graph-PDE Comparisons}\label{subsec:Comparisons of graph problems with continuum limits}

With homogenized continuum-limit parameters, we compare against NEP solutions for the three random-graph types, examining different versions of \eqref{eq:random_continuum_limit}.

\begin{figure*}
    \centering
    \includegraphics[width=0.85\textwidth]{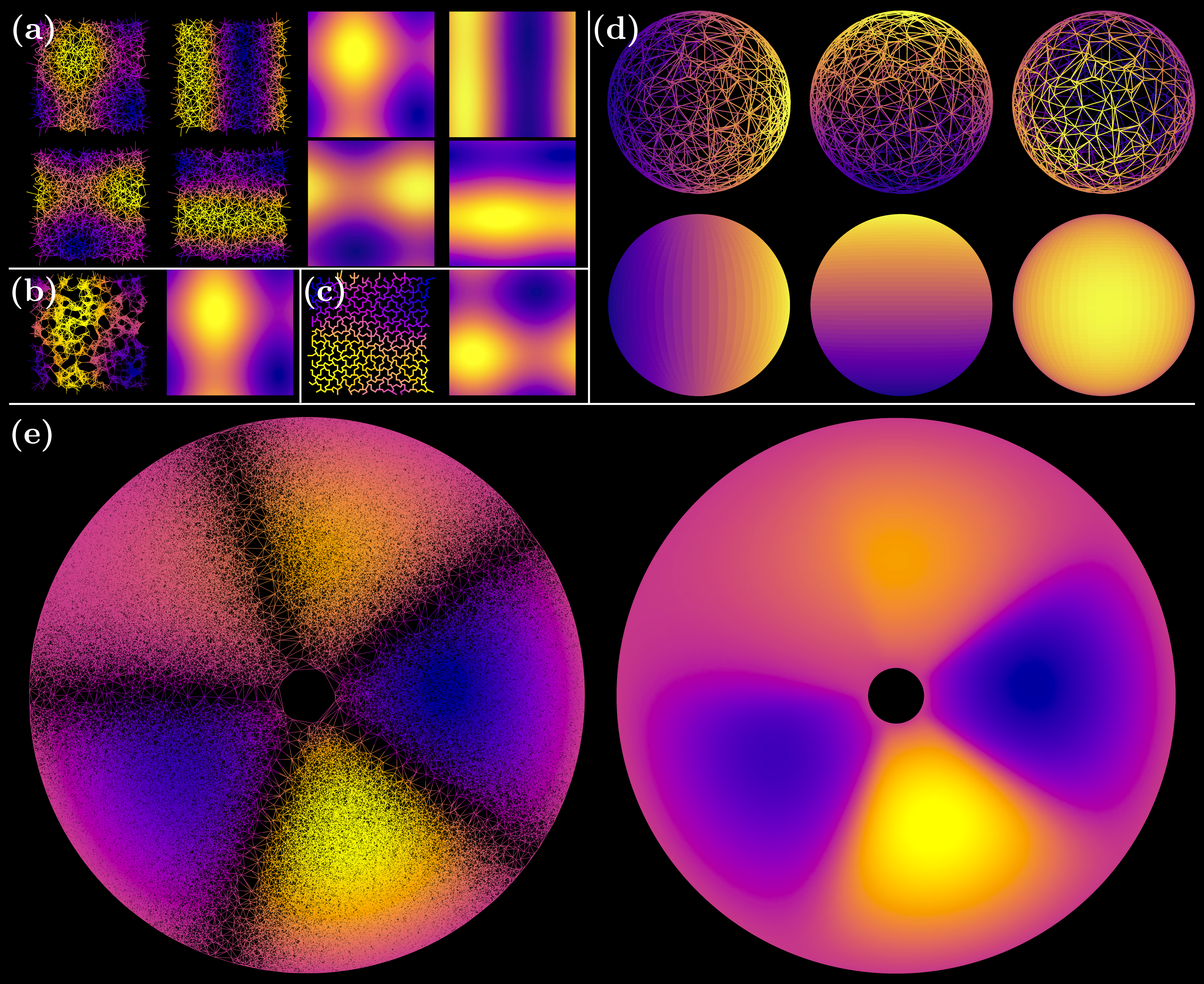}
    \caption{(a)-(c) Visual comparison of graph and PDE modes for a uniformly random Delaunay triangulation, random geometric (neighborhood) graph (RGG), and aperiodic monotile on a unit periodic torus, respectively. The first nontrivial PDE modes are four-fold degenerate. The graph-structure randomness breaks the multiplicity (for any given realization). Each comparison comprises a superposition of $\sin(2\pi x), \cos(2 \pi x)$ and $\sin(2\pi y), \cos(2 \pi y)$, with amplitudes fit to the corresponding graph mode (we only show 1/4 of the modes for the RGG and aperiodic monotile). 
    (d) Visual comparison of the first nontrivial graph and PDE modes for a uniformly random Delaunay triangulation in the 2-sphere. The $j = 1$ graph modes are approximately three-fold degenerate and rotated to match Cartesian solid harmonics, $\propto x, z, y$.
    (e) Illustration of modes on an inhomogeneous random Delaunay triangulation of an annulus. } 
    \label{fig:random}
\end{figure*}

\paragraph{Random graphs in the square flat torus} 
We construct flat square torus graphs the same way as in the construction of the corresponding cell problems.
For each graph type, we solve the NEP on ensembles of increasing density, comparing with the standard solutions of \eqref{eq:random_continuum_limit},
\begin{align}
    \lambda_{m, n}
    & \eeq 
    4 \pi^{2} \, \Diffu \left( m^{2} + n^{2} \right),
    \label{eq:delaunay_torus_eigenvalues}
    \\
    f_{m, n}
    & \, \propto \, 
    e^{2 \pi i (m x + n y)} + \text{c.c.},
    \label{eq:delaunay_torus_eigenfunctions}
\end{align}
where $(m, n) \in \{(0,\pm 1),(\pm 1,0)\}$ for the first nontrivial modes. We use the highest-density numerical approximation to $\Diffu$. For the respective Delaunay triangulation, RGG, and aperiodic monotile, $\Diffu \approx 0.9201/2,\, 0.6781/2,$ and $0.6074/2$ respectively.
We show approximate $\mathcal{O}(\ell)$ convergence of the graph eigensolutions to the PDE eigensolutions in SM \cite{supp}.
For the graph modes, randomness induces a four-way splitting of the eigenvalue and pattern selection.
In figures~\ref{fig:random}(a)-(c), we show the first four split nontrivial graph modes projected onto the PDE modes for the random Delaunay triangulation, and one of these for each of the RGG and aperiodic monotile.

\paragraph{Uniformly random Delaunay triangulation of the 2-sphere} We use the 2-sphere to illustrate the continuum limit on a curved manifold. In this case, we use spherical harmonics for the PDE solutions,
\begin{align}
    \lambda_{j}
    & \eeq \Diffu\,  j (j + 1),
    \\
    f_{j, m}
    & \, \propto \,  Y_{j, m}(\theta, \phi) + \text{c.c.},
\end{align}
where $j \in \mathbb{N}$ and $|m| \leq j$.
Again, we compare eigenmodes corresponding to the first nonzero eigenvalue, \textit{i.e.}, $j = 1$ with $m = -1,0,1$.
We show approximate $\mathcal{O}(\ell)$ convergence of the graph eigensolutions to the PDE eigensolutions in SM \cite{supp}.
Figure~\ref{fig:random}(d) shows the corresponding graph and PDE modes for $|V_{n}| \approx 10^{3}$.

\paragraph{Inhomogeneous random Delaunay triangulation of an annulus}
The final example is a prototype for an empirical graph, \textit{i.e.}, with \textit{a priori} unknown continuum coefficients. While we can still compare the graph modes to the approximate PDE mode, we consider the graph as a single instance and derive pragmatic fits for the density, $\mu$, and homogenized $\mathcal{K}$ and $\mathcal{C}$.
We create a non-uniform graph on a cylindrical annulus $0.1 \leq r \leq 1$ by generating $2\times 10^{5}$ vertices using rejection sampling according to the polar coordinate distribution function $(0.02 + r^{2}) (1.02 + \cos(5 \theta))$ and determine the edges via Delaunay triangulation. We use clamped boundary conditions at $r=0.1,\,1$. We construct macroscale vertex subsets $x$ (\eqref{w to x connection}) by centering square patches on a $16 \times 64$ polar grid--adjusting each patch size to contain approximately $|x| = 300$ vertices; the patches often overlap (see \cite{supp}). The homogenization procedure generates a grid of density values, $\mu(r_{i},\theta_{j})$, capacity function values, $\mathcal{C}(r_{i},\theta_{j})$, and three independent components of the symmetric tensor, $\mathcal{K}(r_{i},\theta_{j})$. We find empirically that $\mathcal{K} \approx \tr{\mathcal{K}}/2 \, I$ where $\tr{\mathcal{K}}$ is a smooth function of $r,\theta$ and the remaining deviatoric components are approximately $1\%$ of the leading order. 
We solve the resulting PDE using the Dedalus spectral solver \cite{burns2020dedalus}.
Figure~\ref{fig:random}(e) shows a visual comparison of a graph mode and its continuum approximation.
We compare the first several eigenvalues and eigenfunctions in \cite{supp}.

\section*{Conclusions}

We present a PDE approach to metric graphs for modeling complex networks of nodes connected via continuous edges. The model is not intended to represent metric graph dynamics exactly in all settings. Rather, we present the examples in this work to show the pragmatic alikeness of the continuum, sometimes with only moderately dense networks. \Eqref{action} could also include additional direct pointwise terms at the vertices, but these require no special treatment in the continuum limit and simply add to the eventual final result.  The model presented here is similar to ``constitutive equation'' approaches in other topics, like Murray's law or Darcy-Brinkman in porous media \cite{Murray1926, Brinkman1949,Weinbaum2003,Holter2017,Chen2022}, and the Q-tensor in liquid-crystal or polymer fluidics modeling \cite{DeGennes1971, Schopohl1987, Wensink2012, Zhang2017, Mackay2020}. The idea for future applications is to start from the PDE standpoint and construct empirically plausible edge tensors $R$ and vertex densities $\mu$. We are particularly intrigued by the possibility of using different $R$, $\mu$ parameters to model different brain regions in complex neuron tissue. Other future work should consider nonlinear effects in dense metric graph material, where the macroscopic parameters could respond dynamically to underlying changes in graph morphology.

\subsubsection*{Materials and methods}
Code for reproducing the numerical results can be found at \url{https://github.com/sidneyholden1/laplace_operator_metric_graph}.

\subsubsection*{Acknowledgements}
Both authors thank Robby Marangell for conversations about the properties of quantum graphs, John Baez for pointing to the truncated square and trihexagonal lattices, and Alexander Morozov for helpful comments on the manuscript.

% \bibliographystyle{apsrev4-2}
% \bibliography{references}

%

\foreach \x in {1,...,15}
{%
\clearpage
\includepdf[pages={\x}]{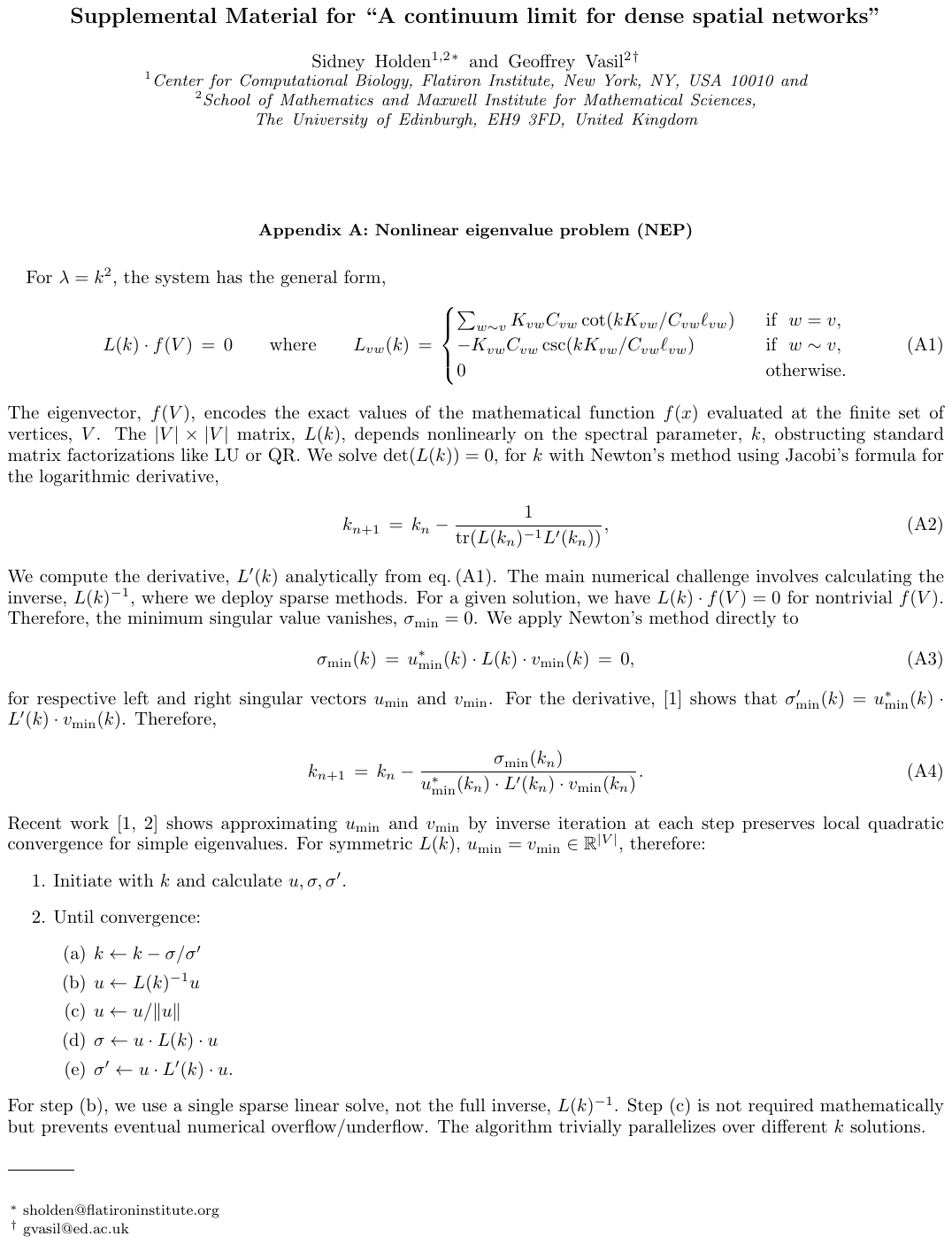} 
}

\end{document}